\documentclass[twocolumn,amsmath,amssymb,superscriptaddress]{revtex4}

\usepackage{graphicx} 
\usepackage{dcolumn}  
\usepackage{bm}       

\begin{document}

\title{Electron correlations and bond-length fluctuations in copper
oxides: from Zhang--Rice singlets to correlation bags}

\author{L. Hozoi}
\affiliation{Max-Planck-Institut f\"{u}r Festk\"{o}rperforschung,
             Heisenbergstrasse 1, 70569 Stuttgart, Germany }

\author{S. Nishimoto}
\affiliation{Institut f\"{u}r Theoretische Physik,
             Universit\"{a}t G\"{o}ttingen, Friedrich-Hund-Platz 1,
             37077 G\"{o}ttingen, Germany }

\author{A. Yamasaki}
\affiliation{Max-Planck-Institut f\"{u}r Festk\"{o}rperforschung,
             Heisenbergstrasse 1, 70569 Stuttgart, Germany }

\date{\today}

\begin{abstract}
We perform first principles, multiconfiguration calculations on
clusters including several CuO$_6$ octahedra and study the
ground-state electron distribution and electron--lattice couplings
when holes are added to the undoped $d^9\,p^6$ configuration.
We find that the so-called Zhang--Rice state on a single CuO$_4$
plaquette is nearly degenerate with a state whose leading
configuration is of the form Cu\,$d^9$--\,O\,$p^5$--\,Cu\,$d^9$.
A strong coupling between the electronic and nuclear motion gives
rise to large inter-site charge transfer effects for half-breathing
displacements of the oxygen ions.
Under the assumption of charge segregation into alternating hole-free
and hole-rich stripes of Goodenough \cite{jbg_02,jbg_03}, our results
seem to support the vibronic mechanism and the traveling
charge-density wave model from Refs.\cite{jbg_02,jbg_03} for the
superconductivity in copper oxides.
\end{abstract}


\maketitle

\section{Introduction}

It is widely believed that the lattice degrees of freedom should play
some role in the high-$T_{c}$ superconductivity in doped copper oxides.
In a recent series of articles, see for example
\cite{jbg_02,jbg_03,jbg_04}, J. B. Goodenough and coworkers predict
for these systems the existence of strong lattice instabilities
associated with two distinguishable (equilibrium) Cu--O bond lengths
and suggest that long range ``ordering'' of bond-length
fluctuations would stabilize some kind of traveling charge-density wave (CDW) in the superconducting state.
Goodenough \textit{et al.} propose a model where the hole--hole
Coulomb repulsion and relatively strong antiferromagnetic (AFM)
$d^9$--\,$d^9$ couplings give rise at $x\!=\!1/6$ doping and low
temperatures to a configuration with alternating, $1/3$-doped and
hole-free, Cu--O--... stripes.
Within the 1/3-doped Cu--O--... rows the holes occupy Cu--O--Cu
bonding orbitals.
The superconductivity is associated in
Refs.\cite{jbg_02,jbg_03,jbg_04} with itinerant heavy vibrons that
originate from the coupling between some particular (dynamic)
ordering of these Cu--O--Cu ``hole bags'' and in-plane phonons.

The present paper is an attempt towards a better understanding of the
nature of the local electron--electron and electron--lattice interactions in copper oxides.
We employ \textit{ab initio} methods from traditional quantum
chemistry. Our calculations are performed on finite clusters
including several CuO$_6$ octahedra. We pay special attention to the
situation where one hole is added to the undoped, formally
Cu\,$d^9$ O\,$p^6$, configuration.
Results from a square 5-octahedra cluster indicate that the so-called
Zhang--Rice (ZR) singlet on a CuO$_4$ plaquette \cite{ZR_86} may
become ``unstable'' with respect to a lower symmetry oxygen-hole
state with a Cu\,$d^9$--\,O\,$p^5$--\,Cu\,$d^9$ leading
configuration.
This is just the 1-hole Cu--O--Cu correlation bag of Goodenough
\cite{jbg_02}.
We further investigate local interactions within a 1-hole,
4-octahedra linear cluster. We find evidence for strongly anharmonic,
Jahn--Teller (JT) type effects for certain Cu--O bond distortions. The oxygen-atom vibrations are coupled with strong inter-site charge
transfer (CT) effects.
Our findings seem to support a mechanism of the type proposed by
Goodenough \textit{et al.} \cite{jbg_02,jbg_03,jbg_04} for the high
temperature superconductivity.

\begin{figure}[b]
\includegraphics[angle=270,width=0.80\columnwidth]{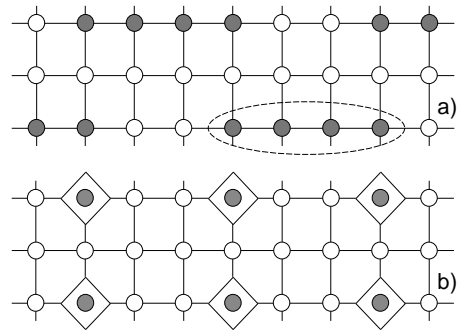}
\caption{(a) 2-(doped-)hole, 4-Cu correlation bags in the CuO plane
at 1/6-doping as suggested by Goodenough \cite{jbg_03}, see text.
Only the Cu sites are shown.
A hole-free ($d^9\,p^6$) chain is drawn with open circles.
(b) Hole-free and 1/3-doped stripes in a Zhang--Rice picture.
The squares represent ZR singlets on CuO$_4$ plaquettes.
A relative shift between neighboring doped stripes should minimize
the Coulomb repulsion. Intuitively, a 2-hole correlation bag forms
by ``clustering'' of two ZR singlets.}
\end{figure}

It was suggested in Ref.\,\cite{jbg_03} that at $x\!=\!1/6$ doping
1-hole Cu--O--Cu units could cluster to form spin-paired 2-hole bags
including four Cu centres. A hypothetical ordered configuration of
such 2-hole bags is shown in Fig.1(a).
In a Zhang--Rice picture, the $1/3$-doped stripes can be represented
as in Fig.1(b), for example.
Our analysis is carried out on 1-hole clusters corresponding to a
somewhat intermediate arrangement, as we shall discuss below.

\section{Local correlations in the CuO plane: CuO$_4$ Zhang--Rice
singlets versus Cu\,$d^9$--\,O\,$p^5$--\,Cu\,$d^9$ configurations}

First, we perform electronic structure calculations on a square
cluster including a central CuO$_6$ octahedron and the four adjacent
in-plane octahedra.
This [Cu$_5$O$_{26}$] cluster is embedded in a matrix of
point charges that reproduce the Madelung potential associated to
the undoped La$_2$CuO$_4$ crystal.
The nearest neighbor cations are represented by effective ion
potentials. We use the tetragonal crystal structure of
La$_{1.85}$Sr$_{0.15}$CuO$_4$ as reported by Cava \textit{et al.}
\cite{LaCuO_cava87}, with in-plane Cu--O distances of 1.89\,\AA .

We rely on an \textit{ab initio}, wave-function-based
multiconfiguration (MC) approach \cite{book_qc}.
Our starting point is a Hartree--Fock self-consistent-field (SCF)
calculation for a closed-shell Cu\,$d^{10}$ O\,$p^6$ configuration
of the [Cu$_5$O$_{26}$] cluster.
We remove then six electrons. We construct thus a multi-determinant
(or multi-configuration) wave-function where ``on top'' of a number
of closed electron shells six electrons/holes are distributed in all
the possible ways over six orbitals. Such a configurational space is
referred to as a complete active space (CAS) \cite{book_qc}.
Those electrons and those orbitals defining the MC space are called
active.
Due to the relatively large size of the cluster we use effective core
potentials (ECPs) for the Cu $1s$,\,$2s$,\,$2p$,\,$3s$ and the four
bridging-oxygen (O$_b^i$, see Fig.2) $1s$ core shells.
The ECP basis sets of Huzinaga \textit{et al.} \cite{ECPs_OCu} were
applied for these ions, with the following contractions:
Cu $9s6p6d/2s2p3d$ and O $5s6p1d/2s3p1d$. For the other oxygens
we applied the basis set of Widmark \textit{et al.} \cite{ANOs_O}
contracted to $2s2p$. All calculations were performed with the
\textsc{molcas} program package \cite{molcas6}.

Unexpectedly, for an undistorted structure with identical Cu--O
distances, the CAS\,SCF ground-state optimization converged to a
broken-symmetry triplet state where the largest weight corresponds
to a Cu$_c$\,$d_{x^2-y^2}^{\,1}$--\,O$_b^1$\,$p_y^1$--\,Cu$_{sq}^1$\,$d_{x^2-y^2}^{\,1}$ configuration (see Fig.2(a)), although
our 5-octahedra cluster has full $D_{4h}$ symmetry \cite{note_1}.
The O$_b^1$ $p_y$ and Cu$_{sq}^1$ $d_{x^2-y^2}$ atomic orbitals
form in this geometry bonding and antibonding combinations.
Contributions from the Cu$_c$ $d_{x^2-y^2}$ and $p_{x}/p_{y}$
orbitals on ligands adjacent to the Cu$_{sq}^1$ cation to this pair
of bonding and antibonding orbitals, while also present, are much
smaller.
Due to correlation, not only the bonding orbital is occupied but a
partial charge transfer into the antibonding orbital occurs.
The latter has an occupation number of 0.27, in comparison to 1.73
for the former.
The formation of such a Cu$_{sq}$\,$d_{x^2-y^2}$--\,O$_b^1$\,$p_y$
bond favors antiferromagnetic $d_{x^2-y^2}$--\,$d_{x^2-y^2}$
interactions over the other three bridging oxygens.

It turns out that the so-called ZR state, where the oxygen hole is
equally distributed over four anions (Fig.2(b)), is approximately 80
meV above the broken-symmetry state with overall triplet spin
multiplicity.
Our data also indicate that it costs 44 meV to couple
ferromagnetically those three Cu$_{sq}$ $d_{x^2-y^2}$ electrons to
the Cu$_{c}$ $d_{x^2-y^2}$ (Fig.2(c)).
Next, we determine the positions of the bridging oxygens O$_b^i$ that
correspond to a minimum of the cluster total energy. According to our results this is the structure where the four Cu$_c$--O$_b^i$ bonds
are all about 5\% shorter. For this arrangement, the ground state is
a ZR like state with full $D_{4h}$ symmetry.
No broken-symmetry solution could be obtained for this geometry.
Now, if we compare Fig.2(b) with Fig.2(c) (or Fig.2(a)) we can
imagine the following scenario:
starting from the minimum-energy structural configuration with
shorter Cu$_c$--O$_b^i$ bonds and a ZR electronic ground state,
(half-)\,breathing displacements of the O$_b$ oxygens may result in a
broken-symmetry state where the leading configuration is of the form
Cu\,$d_{x^2-y^2}^{\,1}$--\,O\,$p_y^1$--\,Cu\,$d_{x^2-y^2}^{\,1}$
and the oxygen $2p$ hole has the largest weight
(\textit{i.e.} is partially transferred, about 0.3 of an electron
charge) onto a single anion.
It is important to mention that the wave functions and the relative
energies for the states discussed above remain largely unchanged when
the active space is extended to include more ligand $2p$ orbitals,
such as the other linear combinations involving O$_b$ $p_y/p_x$
bridging orbitals.

\begin{figure}
\includegraphics[angle=270,width=1.00\columnwidth]{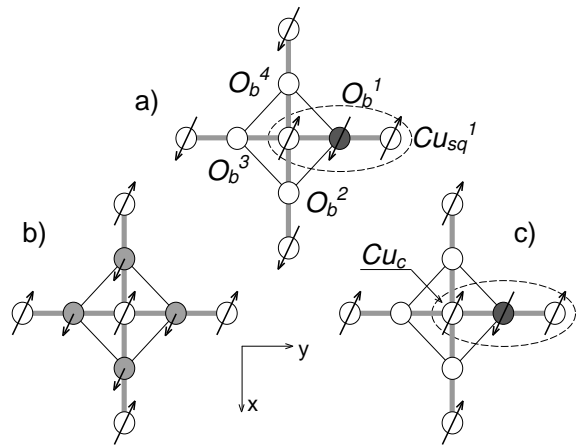}
\caption{(a) Leading configuration for the broken-symmetry,
lowest-energy state in the geometry with identical Cu--O distances,
see text. A CuO$_4$ plaquette and four in-plane Cu neighbors are shown.
(b) Schematic representation of the ZR state on our cluster.
(c) Leading configuration for the lowest excited state in the
undistorted structure. The $d$--$d$ couplings are ferromagnetic. }
\end{figure}

It would be useful to determine how the cluster charge
distribution changes for oxygen-atom distortions on an adjacent
CuO$_4$ plaquette.
However, due to hardware limitations we could not employ sufficiently
accurate basis sets for those oxygens beyond the nearest neighbors
of the central Cu ion.
To study such effects, the ``cheapest'' option is a 1-hole,
2-octahedra cluster.
Calculations on this cluster reveal strong charge fluctuations
between the two symmetric, minimum-energy geometrical configurations
where the in-plane Cu--O distances on only one of the plaquettes are
$5\!-\!6\%$ shorter.
For each of these distorted configurations, the doped hole has the
largest weight on a single O$_4$ plaquette and forms a ZR type
singlet with the Cu $3d$ hole.
In the undistorted geometry, the oxygen hole is mostly localized
on the ligand bridging the two octahedra.
The energy of the high-symmetry state is about 200 meV higher as
compared to the ZR like configurations.
The results indicate strong anharmonic effects, \textit{i.e.} a
one-dimensional double-well, when shifting the bridging oxygen along
the $y$ axis. This in analyzed in the next section.

For an \textit{isolated} Cu$_c$O$_6$ octahedron (or an isolated
5-octahedra square cluster) with no distortions, the simplest way
to view the 1-doped-hole ground-state wave-function is as a
superposition of the four broken-symmetry, degenerate
Cu$_c^{2+}$--\,O$^{-}$ solutions, see for example the discussion
in \cite{nio_bagus99}.
Nevertheless, the calculations on the 2-octahedra cluster show
that there is a finite probability for the ZR singlet to hop to an
adjacent plaquette via such an intermediate broken-symmetry state
with a dominant $d^9$--\,$p^5$--\,$d^9$ configuration.
(Meta)stable $d^9$--\,$p^5$--\,$d^9$ configurations could
actually be induced by longer range Coulomb interactions and AFM
spin correlations in the doped plane.
It seems then that the existence of 1-(doped-)hole Cu--O--Cu entities
that could eventually cluster to form multi-hole correlation bags,
as suggested by Goodenough \cite{jbg_02,jbg_03}, is not an
unrealistic hypothesis.

\section{1-hole, 4-octahedra linear clusters: bond-length fluctuations and inter-site charge transfer}

Inspired by the work of Goodenough \textit{et al.}
\cite{jbg_02,jbg_03,jbg_04} we investigate local many-body
effects and electron--lattice interactions within a 1-(doped-)hole,
4-octahedra [Cu$_4$O$_{21}$] linear cluster.
We apply $C_{2v}$ symmetry restrictions, with $xy$ and $zy$ mirror
planes (see Fig.3), and all-electron basis sets:
Cu $21s15p10d/5s4p3d$ \cite{ANOs_Cu}, O $14s9p4d/4s3p1d$ for the
bridging O$_b$ oxygens and O $14s9p/4s3p$ for the other oxygens \cite{ANOs_O}.

\begin{figure}
\includegraphics[angle=270,width=1.00\columnwidth]{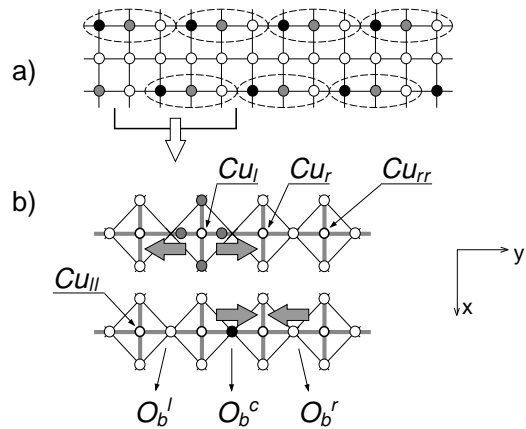}
\caption{(a) Possible charge-ordered configuration with 1-(doped-)hole,
3-Cu spin-singlet ``correlation bags'' along the 1/3-doped stripes,
see text. Only the Cu sites are shown.
A hole-free chain is drawn with open circles.
(b) Four consecutive CuO$_4$ plaquettes along the 1/3-doped chain.
The CAS\,SCF calculations were carried out on such clusters.
The O $2p$ hole is represented with black or gray circles.
For certain phases of the O-atom half-breathing vibrations the
hole can hop along the chain.
From left to right, the Cu and the bridging O atoms along the CuO chain
are labeled as Cu$_{ll}$, O$_b^l$, Cu$_l$, O$_b^c$, Cu$_r$,
O$_b^r$, and Cu$_{rr}$.}
\end{figure}

\begin{table*}
\caption{Contributions of various oxygen $2p$ and metal $3d$ atomic
orbitals to the relevant Cu--O bonding (B) and antibonding (AB)
active (natural) orbitals \cite{note_3}.
In each case, the other three active orbitals have essentially
Cu $d_{x^2-y^2}$ character and occupations of nearly 1\,.
Coefficients smaller than 0.10 are not shown. Natural-orbital
occupation numbers and Mulliken populations (MPs) are also given.
O$_b$ are bridging oxygens along the CuO chain, the O$_a$ ions form
parallel, adjacent in-plain O chains, see Fig.3(b).}
\begin{ruledtabular}
\begin{tabular}{lrrlrrlrrl}
  State \tablenotemark[1]:
&\multicolumn{3}{l}{O$_b^l$-hole, undist. cluster}
&\multicolumn{3}{l}{Cu$_l$O$_4$ ZR, dist. Cu$_l$--\,O$_b$ bonds
                    \tablenotemark[2]}
&\multicolumn{3}{l}{O$_b^c$-hole, undist. cluster}        \\
Orbitals:                &B         &AB          &MPs
                         &B         &AB          &MPs
                         &B         &AB          &MPs           \\
\colrule
O$_b^{ll}$ $p_y$         &--         &0.25       &1.74
                         &--         &--         &1.83
                         &--         &--         &1.84          \\
O$_a^{ll}$ $p_x$ (x2)    &$\mp$0.13  &$\pm$0.26  &1.70
                         &--         &--         &1.83
                         &--         &--         &1.85          \\
Cu$_{ll}$ $d_{x^2-y^2}$  &0.62       &0.72       &1.09
                         &--         &--         &1.16
                         &--         &--         &1.16          \\
O$_b^l$ $p_y$            &0.66       &--0.64     &\textbf{1.27}
                         &0.38       &0.42       &\textbf{1.62}
                         &--         &0.27       &1.77          \\

Cu$_l$  $d_{x^2-y^2}$    &--0.18     &--0.24     &1.15
                         &--0.63     &0.79       &1.01
                         &--0.62     &0.72       &1.08          \\
O$_a^l$ $p_x$ (x2)       &--         &$\mp$0.11  &1.80
                         &$\pm$0.33  &$\pm$0.36  &\textbf{1.64}
                         &$\pm$0.13  &$\pm$0.26  &1.71          \\
O$_b^c$ $p_y$            &--         &--         &1.83
                         &--0.38     &--0.42     &\textbf{1.62}
                         &--0.66     &--0.64     &\textbf{1.27} \\

Cu$_r$  $d_{x^2-y^2}$    &--         &--         &1.15
                         &--         &--         &1.15
                         &0.18       &--0.25     &1.15          \\
O$_a^r$ $p_x$ (x2)       &--         &--         &1.85
                         &--         &--         &1.83
                         &--         &$\mp$0.11  &1.80          \\
O$_b^r$ $p_y$            &--         &--         &1.84
                         &--         &--         &1.84
                         &--         &--         &1.83          \\
\\
Occ. No.                 &1.73       &1.28       &
                         &1.87       &0.13       &
                         &1.72       &0.28       &              \\
\end{tabular}
\end{ruledtabular}
\tablenotetext[1]{ \ The fact that the doped holes have the largest
weight on O $p_x/p_y$ orbitals pointing to the Cu sites
was also found in previous \textit{ab initio} wave-function-based
calculations \cite{cuo_stollhoff,cuo_RLmartin,cuo_calzado00}.
Tendencies towards localization and broken-symmetry solutions were
discussed in \cite{cuo_RLmartin}.}
\tablenotetext[2]{ \ The chain Cu$_l$--O$_b$ bonds are 6\% shorter.}
\end{table*}

\begin{figure}
\includegraphics[angle=0,width=1.00\columnwidth]{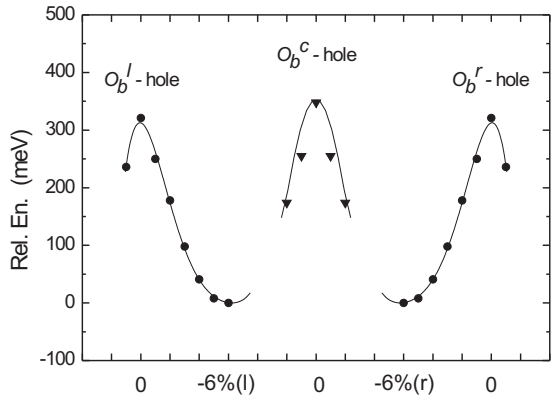}
\caption{Quasi-localized O-hole states on a 1-doped-hole, 4-octahedra
linear cluster, see text.
The energy maxima correspond to states with dominant O$_b$-hole
character in the undistorted structure. The two minima are related
to ZR like states where the Cu--O$_b$ distances on the Cu$_l$O$_4$
plaquette (the left hand minimum) or the Cu--O$_b$ distances on the
Cu$_r$O$_4$ plaquette (the right hand minimum) are 6\% shorter.
From each minimum-energy structural configuration, the two O$_b$
ligands are shifted back to the undistorted geometry in steps of
1\% of the high-symmetry Cu--O bond length.
The energies of these states are shown with black dots.
The lines are a guide for the eye. The last point on each of these
curves corresponds to a configuration where the Cu--O$_b$ distances
on the Cu$_{ll}$O$_4$/Cu$_{rr}$O$_4$ plaquette are shorter, by 1\%.
Other calculations were carried out for the case where the Cu--O$_b$
bonds on the Cu$_l$O$_4$ and Cu$_r$O$_4$ plaquettes are gradually
shortened starting from the undistorted structure and using the
orbitals of the O$_b^c$-hole state as starting orbitals.
Due to convergence problems, only few points could be obtained.
These points are shown as black triangles.}
\end{figure}

Starting from an undistorted structure with identical Cu--O distances,
we perform a number of test calculations to identify the geometry that
minimizes the cluster total energy.
However, only the chain Cu and O$_b$ ions are allowed to move, along
the $y$ axis.
We do not modify the positions of the in-plane oxygens adjacent to
the CuO chain, O$_{a\,}^i$, or the positions of the apical ligands.
We employ a 5\,electron\,/\,5\,orbital CAS, with five electrons
being removed from the closed-shell Cu\,$d^{10}$ O\,$p^6$ configuration.
It is found that the lowest-energy geometry corresponds to a distorted
configuration like that shown in Fig.3(b). The O$_b^c$ atom
is shifted to the left by 6\% of the high-symmetry Cu--O bond
length and the O$_b^l$ atom is shifted by 6\% to the right.
For this geometry an oxygen $2p$ hole/electron is bound to the
Cu$_l$ $d_{x^2-y^2}$ hole/electron into a ZR like singlet state.
We give in Table\,I the composition of the bonding (B)
and antibonding (AB) active (natural) orbitals \cite{note_3} on the
Cu$_l$O$_4$ plaquette, their occupation numbers and the Mulliken
populations (MPs) for the relevant O $2p$ and Cu $3d$ atomic orbitals.
The other three active orbitals have essentially Cu$_{ll\,}$,
Cu$_r$ and Cu$_{rr}$ $d_{x^2-y^2}$ character and occupation
numbers of nearly 1\,.
The overall spin multiplicity is doublet, with the Cu$_r$ and
Cu$_{rr}$ $3d$ electrons coupled antiferromagnetically.

Now, from this geometrical configuration we shift the O$_b^l$ and
O$_b^c$ nuclei back to the high-symmetry, undistorted arrangement in
steps of 1\% of the Cu--O bond length.
The orbitals from the previous step are used each time as input
orbitals \cite{note_4}.
For the undistorted geometry the calculation converges to a state
where the doped hole is mainly localized on the bridging O$_{b\,}^l$
oxygen. Details regarding the composition and the occupation of
the relevant orbitals are given in Table\,I.
It was found that the energy of this O$_{b\,}^l$-hole state in the
undistorted geometry is 320 meV higher then for the ZR like state
in the configuration where the Cu--O$_b^l$ and Cu--O$_b^c$ bonds are
6\% shorter.
In agreement with the results from the previous section, between
these two configurations a transfer of about one third of an
electron/hole occurs from the O$_b^l$ $p_y$ to the O$_a^l$ $p_{x\,}$,
O$_{b}^c$ $p_{y\,}$, and Cu$_l$ $d_{x^2-y^2}$ orbitals, see Table\,I.
The total cluster energy is plotted for different positions of the
O$_{b}^l$ and O$_b^c$ oxygens in the left part of Fig.4\,.

For systems with competing valence configurations like our hole-doped
cluster, different sets of input orbitals can lead to different
solutions.
We performed calculations where we forced the localization of the
doped hole on the central O$_{b}^c$ $p_{y}$ orbital, by adding
some extra positive charge within our point charge embedding.
Using this output as a new set of starting orbitals for CAS\,SCF
calculations with the initial embedding and no distortions, we were
able to obtain a solution with a dominant contribution from the
...--Cu$_l$\,$d_{x^2-y^2}^{\,1}$--\,O$_b^c$\,$p_y^1$--\,Cu$_r$\,$d_{x^2-y^2}^{\,1}$--... configuration.
The energy of this state is 350 meV above the minimum corresponding
to the ZR state on the distorted Cu$_l$O$_4$ plaquette (see Fig.4).
For the O$_{b}^l$-hole state, the
Cu$_r$\,$d_{x^2-y^2}^{\,1}$--\,Cu$_{rr}$\,$d_{x^2-y^2}^{\,1}$
AFM interactions are not perturbed by the presence of the oxygen
hole and its relative energy is somewhat lower, 320 meV.
The fact that the solution with the doped hole mainly localized on
the O$_b^c$ ligand is less accessible can be understood on the
basis of this energy difference between the two O$_b$-hole states.
Regarding the data in Table\,I and Fig.4 we also note that the
O$_b^c$-hole doublet wave-function calculated for the undistorted
geometry misses full $D_{2h}$ symmetry.
Attempts to converge to a fully symmetric wave-function were
unsuccessful.

The structural configuration with 6\% shorter chain Cu$_l$--O bonds
is symmetry equivalent to the situation where the chain Cu$_r$--O
bonds are 6\% shorter.
These two distorted configurations represent the minima of a
one-dimensional double-well potential.
From one minimum to the other, the sum of the Mulliken populations of
the O$_b$ $p_{y\,}$, O$_a$ $p_{x\,}$, and Cu $d_{x^2-y^2}$ orbitals
on each of the Cu$_l$O$_4$ and Cu$_r$O$_4$ plaquettes changes by
about $0.7\,e$.
The total Mulliken electronic charge on each of these plaquettes
changes between the two minima by about $0.5\,e$ (not shown in
Table\,I) \cite{note_6}.
It is well known that the MPs are basis set dependent. Still,
variations in the charge population of a whole CuO$_4$ plaquette
(or CuO$_6$ octahedron) is a reliable quantitative indicator.
An illustrative result is also the variation of the total Mulliken
electronic charge for the Cu$_l$O$_6$ and Cu$_{ll}$O$_6$ octahedra
when going from the state with a
Cu$_{ll}$\,$d_{x^2-y^2}^{\,1}$--\,O$_b^l$\,$p_y^1$--\,Cu$_l$\,$d_{x^2-y^2}^{\,1}$ leading configuration to the valence structure where
the hole is mainly on the O$_{b}^r$ oxygen. This value is $0.95\,e$.
About the same amount of charge, one hole/electron, should be
transferred along the chain between ``traveling'' next nearest
neighbor ZR singlets.

As mentioned in the previous section, we performed extra calculations
on a 1-hole, 2-octahedra cluster.
The doped hole is confined to a smaller region in space in this
situation.
In addition, charge relaxation effects within the neighboring
chain octahedra can not be accounted for.
Therefore, the CuO$_4$--CuO$_4$ CT is weaker when distorting the
Cu—O bonds, about $0.3\,e$.
For this smaller cluster it is possible however to use a more
flexible active space, including all O $2p$ and Cu $3d$ orbitals.
This corresponds to distributing 3 holes over 43 orbitals.
With such an active space we found a CuO$_4$--CuO$_4$ CT of $0.4\,e$
for the half-breathing distortions.

\begin{figure}
\includegraphics[angle=270,width=1.00\columnwidth]{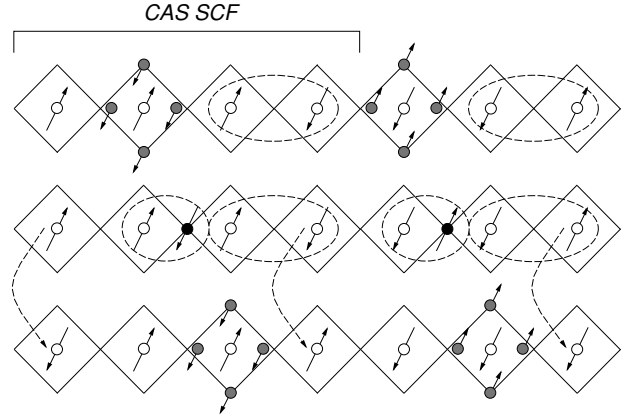}
\caption{ Possible representation of the traveling
charge-density/spin-density wave along a 1/3-doped
CuO$_4$--CuO$_4$--... chain.
The O $2p$ holes are shown with black or gray circles.
For each sequence, the spins on the first four plaquettes display
schematically \cite{note_7} the CAS\,SCF results.
The ellipses indicate spin-singlet couplings.
Dashed arrows indicate a ``spin-flip'' process.}
\end{figure}

The results reported here are quite remarkable.
They show that within a 1-doped-hole cluster including several
CuO$_6$ octahedra, complex electron correlations give rise to
quasi-localized O $2p$ hole states and strong charge inhomogeneity.
Several competing O-hole valence configurations were identified.
The near-degeneracy between different hole configurations and the
electron--lattice interactions cause strongly anharmonic, pseudo JT
effects and facilitate the hopping of the charge carriers.
Starting from a minimum-energy geometry where shorter ``chain''
Cu--O bonds on a CuO$_4$ plaquette give rise to a quasi-localized ZR
like state, a half-breathing motion of the chain oxygens with a
phase difference of $\pi$ between adjacent plaquettes (see Fig.3(b))
leads to a CuO$_4$--\,CuO$_4$ charge transfer of about $0.5\,e$ over
an energy barrier of 350 meV (175 meV per chain oxygen).
The minima of the one-dimensional double-well imply in-line
O-atom displacements about the middle positions with amplitudes of
only 0.11\,\AA \ ($\pm6\%$ of the Cu--O bond length).
Although we did not investigate the effect of further Cu--O$_b$
distortions, it seems that a whole electron charge moves along the
chain when a ...--CuO$_4$--... \cite{note_7} ZR like singlet is
``transferred'' to a next nearest neighbor plaquette.

It is possible that at about 1/6-doping an ordered arrangement of
hole-free and 1/3-doped stripes is formed within the CuO plane
\cite{note_2}.
Under this assumption our 1-hole, 4-octahedra cluster is related to
the unit cell of the charge-ordered 1/3-doped stripe shown in
Fig.3(a) (such a ``unit cell'' actually contains only three in-line
octahedra).
We speculate then that longer range hole--hole Coulomb interactions
associated with cooperative, correlated oxygen-atom displacements
like those described in the previous paragraph along a chain of
several
Cu$_{ll}$--\,O$_b^l$--\,Cu$_l$--\,O$_b^c$--\,Cu$_r$--\,O$_b^r$
units should lower the energy barrier corresponding to the
inter-site hole hopping and make possible an ordered, collective
motion of the doped holes.
Under the assumption of an alternating 1/3-doped/hole-free stripe
configuration, our results seem to support thus a vibronic mechanism
like proposed by Goodenough \cite{jbg_02,jbg_03} for the
superconductivity in cuprates.
It still needs to be investigated whether (and why) nearest neighbor
1-hole correlation bags would ``cluster'' along the chain to form a
(traveling) CDW with spin-paired 2-hole, 4-Cu units
separated by two antiferromagnetically-coupled Cu $d^9$ ions as
suggested in Ref.\,\cite{jbg_03} (see Fig.1(a)).
The longer range interactions responsible for the formation of such
paired oxygen holes should be however weaker than those causing the
local JT/CT effects we illustrate in Fig.4 and Table\,I. They
probably merely modify the amplitudes of the O-atom displacements
that determine the double-well potential.
It should be pointed out at this stage that the CAS\,SCF results
suggest that nearest neighbor oxygen holes have already antiparallel
spins, see Fig.5\,.

Obviously, reliable estimates for the strength and the effects of the
longer range couplings, intra- and inter-chain, are required for
quantitative predictions.
We note that in this model the effect of the inter-chain Coulomb
interactions should be stronger for CuO multilayer structures like
in the mercury compounds and might contribute to a higher $T_c$\,.
The critical temperature $T_c$ should be related to thermal
fluctuations that disrupt the phase coherence among the kind of
O-atom displacements described above.
At temperatures well above $T_c$ the striped arrangement is
probably destroyed.
For a certain temperature interval above $T_c$, the transition from
a charge-inhomogeneous phase with parallel chains (or
fragments of chains) of 1-hole, 3-octahedra units but
``uncorrelated'' bond-length fluctuations to
a totally disordered charge configuration
could provide an explanation for the $(\pi,\pi)-(\pi,\,0)$
transfer of spectral weight in the angle-resolved photoemission
spectroscopy (ARPES) data \cite{cuo_arpes}.

Our [Cu$_4$O$_{21}$] linear cluster seems to be a reasonable
choice for studying local correlations for a 1/3-doped chain of
octahedra. The quasi-localized character of the oxygen hole (see
Table\,I) justifies such a material model. This cluster model
would probably be inappropriate for the 1/2-doped chain (or
stripe) in La$_{1.48}$Nd$_{0.40}$Sr$_{0.12}$CuO$_4$
\cite{stripes_tranquada95}, where the doped-hole orbitals have
much larger overlap. Also, although the CAS\,SCF method can not
yield highly accurate energies, such an approach is sufficient for
an accurate description of the electron distribution provided the
proper active space is chosen. The near-degeneracy and CT effects
illustrated by our CAS\,SCF results should be then a genuine
characteristic of the CuO$_4$--CuO$_4$--... chain. On the other
hand, the CAS\,SCF data normally gives a rather qualitative (or
semi-quantitative) description for potential-energy landscapes
like that associated with the O-atom half-breathing displacements.
A more elaborate treatment, \textit{e.g.} multi-reference
configuration-interaction or multiconfiguration perturbation
theory \cite{book_qc}, is needed for calculating accurate
potential curves and surfaces. Such calculations will be
documented in an upcoming publication.

A matter of concern with our calculations may be the fact that we
used experimental structural data. It would be desirable to
start, for example, from a fully optimized structural configuration
of the undoped system. For such an optimized geometry, the in-plane
Cu--Cu distance would be of particular interest.
It is known that the \textit{ab initio} periodic Hartree--Fock (HF)
calculations in the linear-combination of atomic orbitals (LCAO)
formulation slightly overestimate the lattice parameters in
transition metal oxides, for example by 2\% in MnO and NiO
\cite{geopt_mno_nio} and almost 4\% in Cu$_2$O \cite{geopt_cu2o}.
To reduce these deviations, post-HF techniques are needed
\cite{note_5}.
Nevertheless, structure optimizations for complex periodic systems
like La$_2$CuO$_4$ at a correlated, post-HF level are not possible
yet.
Intuitively, we expect that for slightly larger Cu--Cu distances in
our cluster the doped hole is somewhat more localized on the
distorted CuO$_4$ plaquette (see Table\,I).
In this case the half-breathing O-atom vibrations should imply more
CT between adjacent plaquettes but over a higher energy barrier.

\section{Conclusions}

We study local electron correlation effects for hole-doped CuO clusters
including several CuO$_6$ octahedra. We use the structural data
reported for the tetragonal lattice of La$_{1.85}$Sr$_{0.15}$CuO$_4$.
Results of \textit{ab initio} MC\,SCF calculations on a 1-hole,
4-octahedra linear cluster indicate a double-well potential for
half-breathing O-atom displacements along the O--Cu--O--Cu--O row.
The Cu--O bond-length fluctuations are coupled with large inter-site
charge transfer effects.
For distortions of $\pm6\%$ of the Cu--O bond length, charge
fluctuations of about $0.5\,e$ are observed between nearest neighbor
CuO$_4$ plaquettes.
It seems also that a whole electron charge would move along the chain
when a ...--CuO$_4$--... ZR like singlet is transferred to a next
nearest neighbor plaquette.

For a 1/6-doped CuO plane we adopt the hypothesis of charge segregation
\cite{note_2} into alternating 1/3-doped and hole-free Cu--O--...
stripes of Goodenough \textit{et al.} \cite{jbg_02,jbg_03}.
Under the charge segregation hypothesis, our findings seem to
support the vibronic mechanism and the traveling
charge-density/spin-density wave (CDW/SDW) model of Goodenough for
the high-$T_c$ superconductivity in cuprates.
In this scenario the superconducting state would imply long range ``correlation'' (or phase coherence) among the type of O-atom
displacements shown in Fig.3(b).
We arrive in the end to the kind of picture proposed by Egami
\textit{et al.} \cite{egami_00}\,: electron correlation --- according
to our calculations, the near-degeneracy between the quasi-localized
ZR like state on a CuO$_4$ plaquette and a state whose leading
configuration is of the form Cu\,$d^9$--\,O\,$p^5$--\,Cu\,$d^9$
plus the longer range hole--hole Coulomb repulsion --- is the
engine, while the driver is the lattice vibration.

Work combining results of first principles MC\,SCF calculations on
small clusters and results of calculations for a Hubbard type model Hamiltonian on extended CuO$_4$--CuO$_4$--... chains is in progress.
It is hoped that such investigations will enable reasonable estimates
of the longer range interactions and other (quantitative) predictions.
In a recent study combining the two approaches we were able to give a
rather realistic description of the phase transition in the
mixed-valence NaV$_2$O$_5$ compound \cite{navo}.
``Correlated'' calculations for a bi-dimensional Cu--O structure to
investigate the charge segregation scenario in the CuO plane at
1/6-doping are unfeasible at this moment.

The JT/CT effects illustrated by our \textit{ab initio} results are
much too robust to be taken just as an artifact
of our finite clusters or/and of other approximations.
We believe that these results define a few basic requirements that a
model Hamiltonian approach should meet for realistic predictions in
the optimally doped and under-doped copper oxide compounds.

\section{Acknowledgements}

We thank O. K. Andersen and O. Jepsen for encouraging this study and
A. T. Filip, O. Gunnarsson, and G. Stollhoff for fruitful discussions. L. H. acknowledges financial support from the
Alexander von Humboldt Foundation.


\begin{references}
\bibitem{note_1} The calculations were performed with $C_h$ symmetry
restrictions, with the $xy$ plane as a mirror plane.
\bibitem{note_3} The natural orbitals are obtained such as to
diagonalize the one-electron density matrix.
\bibitem{note_4} A set of input orbitals must be provided for the
CAS\,SCF calculation. These input orbitals can either be from a
Hartree--Fock closed-shell (Cu\,$d^{10}$\,O\,$p^6$) calculation or
from a previous CAS\,SCF calculation on the same system.
\bibitem{note_7} The O $2p$ hole also has some weight on
farther anions, see Table\,I.
\bibitem{note_2} There is much experimental data that suggest the
formation of such stripes, see for example the discussions in
\cite{jbg_02,jbg_03}, \cite{egami_00} and references therein.
\bibitem{note_5} Algorithms able to treat electron correlation
explicitly in periodic systems are being developed, see for example
\cite{cuo_stollhoff,postHF_crystals}.
\end{references}
\end{document}